\documentclass{IEEEcsmag}

\usepackage[colorlinks,urlcolor=blue,linkcolor=blue,citecolor=blue]{hyperref}
\expandafter\def\expandafter\UrlBreaks\expandafter{\UrlBreaks\do\/\do\*\do\-\do\~\do\'\do\"\do\-}
\usepackage{upmath,xcolor}
\usepackage{orcidlink}
\jvol{XX}
\jnum{XX}
\paper{8}
\jmonth{}
\pubyear{2026}
\usepackage{url}

\usepackage{array}
\usepackage{booktabs}
\usepackage{tabularx}
\usepackage{graphicx}

\begin{document}

\title{Heterogeneous Space Data Centers: A Key Paradigm for Future Non-Terrestrial and In-Orbit Services}

\author{Peng Hu*\orcidlink{0000-0002-9069-0484}, Mustafa Alhassan\orcidlink{0009-0002-1356-4772}, Amjad Iqbal\orcidlink{0009-0009-3614-1554}}
\affil{Department of Electrical and Computer Engineering, University of Manitoba, Winnipeg, MB, R3T 5V6, Canada}

\markboth{}{}

\begin{abstract}\looseness-1Lower space-launch cost and advances in AI are driving space-based data centers (SDCs) that provide AI-centric computing and communication services to in-orbit and non-terrestrial assets. However, architectural principles, service models, and energy provisioning for such SDCs, particularly in low-Earth orbit (LEO), remain largely unexplored. We introduce heterogeneous SDCs composed of large and small nodes, differentiated by computing capability, size, and power budget, to support applications spanning ground, aerial, and in-orbit users. We define use cases, node sizing and placement strategies, and interconnection topologies integrating space-to-space and space-to-ground links. Focusing on dawn-dusk LEO configurations, we analyze solar power availability, visibility windows, and latency-aware coverage. Using baseline workloads for typical scenarios, we evaluate latency and service reliability. We quantify energy requirements of different SDC roles and deployment scales, highlighting trade-offs among node size, orbit selection, and service portfolio, and provide design guidance and future research directions for AI infrastructure in space.
\end{abstract}

\maketitle
\chapteri{O}rbital artificial artificial intelligence (AI) infrastructure has attracted growing interest from industry and research communities. For example, on August 04, 2026, SpaceX and NVIDIA jointly announced the Starmind AI1 satellite, and SpaceX's regulatory filings indicate an eventual constellation of up to one million such nodes \cite{spacex}. Google proposed Suncatcher, a constellation of solar-powered satellites equipped with Tensor Processing Unit (TPU) accelerators, motivated by the finding that photovoltaic arrays in a dawn–dusk sun-synchronous orbit (SSO) can generate substantially more power than comparable terrestrial installations \cite{google}. Axiom Space, Starcloud, and China's Three-Body Computing Constellation are pursuing comparable architectures \cite{Axiom,bloch2026earth}, building on Hewlett Packard Enterprise's Spaceborne Computer-2, which has hosted edge-computing and AI experiments aboard the International Space Station since 2021.

These developments collectively signal lower-cost access to space, driven by reusable launch vehicles and advances in physical AI, enabling a new class of infrastructure, including space-based data centers (SDCs). SDCs could provide AI-centric computing and communication services in space for both space-based and terrestrial users. However, existing literature and industry proposals lack architectural principles, service models, and energy-provisioning strategies for SDCs, and most assume a single, largely homogeneous satellite class. This leaves several architectural questions open as constellations scale.

First, it is unclear whether all nodes need full-scale AI training power and thermal capacity, or whether lower-power nodes could handle latency-sensitive, in-situ workloads while fewer high-capacity nodes handle aggregation and heavier computation. Recent analysis suggests that training frontier-scale models in orbit is unlikely to be cost-competitive with terrestrial facilities in the near term, whereas inference appears more viable \cite{van2026cost}, which favors a differentiated rather than uniform node architecture. Second, the interconnection strategy among SDC nodes, ground infrastructure, and the aerial layer of non-terrestrial networks (NTN), such as high-altitude platform stations (HAPS) \cite{geraci2022integrating}, remains undefined. Third, orbit selection strongly affects power availability, visibility, and latency. Independent proposals have converged on the dawn–dusk SSO, which avoids Earth's shadow and provides near-continuous solar illumination, and this convergence warrants systematic analysis.

In this article, we introduce a heterogeneous SDC (HSDC) architecture comprising small SDC (S-SDC) and large SDC (L-SDC) nodes, designed to support ground, aerial, and in-orbit users. The remainder of the article explains why SDCs require heterogeneous nodes, outlines use cases, evaluates the architecture through simulation, and concludes with design guidance and research directions.

\section{SDC VS HETEROGENEOUS SDC}\label{sec:sdc-hsdc}
The terms AI DC, orbital DCs, and DCs in space, collectively referred to as SDCs, describe in-orbit platforms ranging from single-accelerator payloads on Earth-observation (EO) satellites to constellation-scale proposals such as Starmind. An SDC is an in-orbit platform that performs substantive computation, such as training, inference, storage, or data processing, rather than being limited to sensing or communication relay. Under this definition, HPE Spaceborne Computer-2 \cite{Hewlett2021}, the NVIDIA H100-class payload flown by Starcloud \cite{bloch2026earth}, and TPU clusters under development for Project Suncatcher \cite{google} all qualify as SDCs, despite substantial differences in scale and purpose.

Existing SDC designs assume largely interchangeable satellites, each with a comparable compute and power envelope, differing mainly in assigned workload rather than in structural capabilities. This assumption is hard to justify when a constellation must serve heterogeneous users: a small satellite generating a burst of imagery, a HAPS platform relaying connectivity to an underserved region, and a large in-orbit training job differ substantially in latency tolerance, data volume, and power requirements. Provisioning every node for the most demanding case is costly, and orbital systems can least afford to waste mass and power.

\begin{table*}[t]
\centering
\caption{Computing accelerator platforms for HSDC nodes} 
\label{tab:gpu}
\footnotesize                      
\renewcommand{\arraystretch}{1.25} 
\setlength{\tabcolsep}{2pt}
\begin{tabular}{@{}l >{\raggedright\arraybackslash}p{3.6cm}
                  >{\raggedright\arraybackslash}p{3.2cm}
                  >{\raggedright\arraybackslash}p{4.4cm}
                  >{\raggedright\arraybackslash}p{4.2cm}@{}}
\hline
\textbf{} & \textbf{Accelerator Device} & \textbf{Platform} & \textbf{Status} &
\textbf{Power Budget} \\ \hline

1.  & NVIDIA Vera Rubin NVL72 & Starmind AI1 & planned (Q4 2027) & Roughly 180--220\,kW per rack  \\
2.  & NVIDIA GB300 NVL72 & Blackwell Ultra (rack) & Considered for AI1 design; superseded by Vera Rubin & About 120--140\,kW per rack \\
3.  & Google Trillium TPU (4 chips) & Suncatcher MVP & launched (Oct 1, 2026) & About 1\,kW satellite power \\
4.  & NVIDIA H100 & Starcloud-1 & launched (Nov 2, 2025) & Up to 700\,W (SXM); 350--400\,W (PCIe/NVL) \\
5.  & NVIDIA H100 + Blackwell B200 & Starcloud-2 & planned (2027) & About 8\,kW compute satellite  \\
6.  & NVIDIA T4 & HPE Spaceborne Computer-2 & launched (Feb 20, 2021) & 70\,W maximum \\
7.  & NVIDIA Jetson AGX Thor & Aethero NxN/NxA  & space computers in development, no flight yet & Configurable 40--130\,W \\
8. & NVIDIA Jetson AGX Orin & Loft Orbital YAM-9 (JPL NAVI-Orbital) & launched (Nov 28, 2025) & Configurable 15--60\,W \\
\hline
\end{tabular}
\end{table*}
This motivates the proposed HSDC paradigm, in which S-SDC and L-SDC nodes differ primarily in computing capability, size, and power budget while operating cooperatively. An S-SDC is a lightweight, low-power node that serves nearby traffic with minimal latency, whereas an L-SDC is a larger, higher-power node that aggregates demand across S-SDCs and provides capacity beyond that of individual S-SDCs.

Following the HSDC paradigm, S-SDC nodes are expected to support latency-sensitive, in-situ workloads, including real-time computing for nearby spacecraft, inference offloading (such as processing high-resolution satellite imagery), localized storage, and computation for space situational awareness (SSA) and space traffic management. In these domains, AI-based conjunction assessment and maneuver planning are increasingly valuable as low-Earth orbit (LEO) populations grow rapidly \cite{kua2021internet}. S-SDCs can also host digital twin (DT) services that maintain a near-real-time representation of the spacecraft state and support early-stage in-orbit manufacturing. By contrast, L-SDC nodes handle workloads that exceed the capacity of individual S-SDCs, including large-scale training, cross-constellation aggregation, and computing services for the broader NTN layer, such as HAPS operating at approximately 20 km altitude to connect unserved or hard-to-reach regions \cite{geraci2022integrating}.

In summary, the two node classes have different power requirements. An S-SDC can be powered by its own solar array and battery within the size constraints of current large LEO satellite buses, whereas an L-SDC requires power and thermal subsystems comparable to those of the Starmind AI1 platform~\cite{spacex}. Table~\ref{tab:gpu} lists computing accelerators that have flown or have been announced for SDCs, along with their power budgets.

\section{HSDC ARCHITECTURE}\label{sec:arch}
\begin{figure*}
\centerline{\includegraphics[width=32pc]{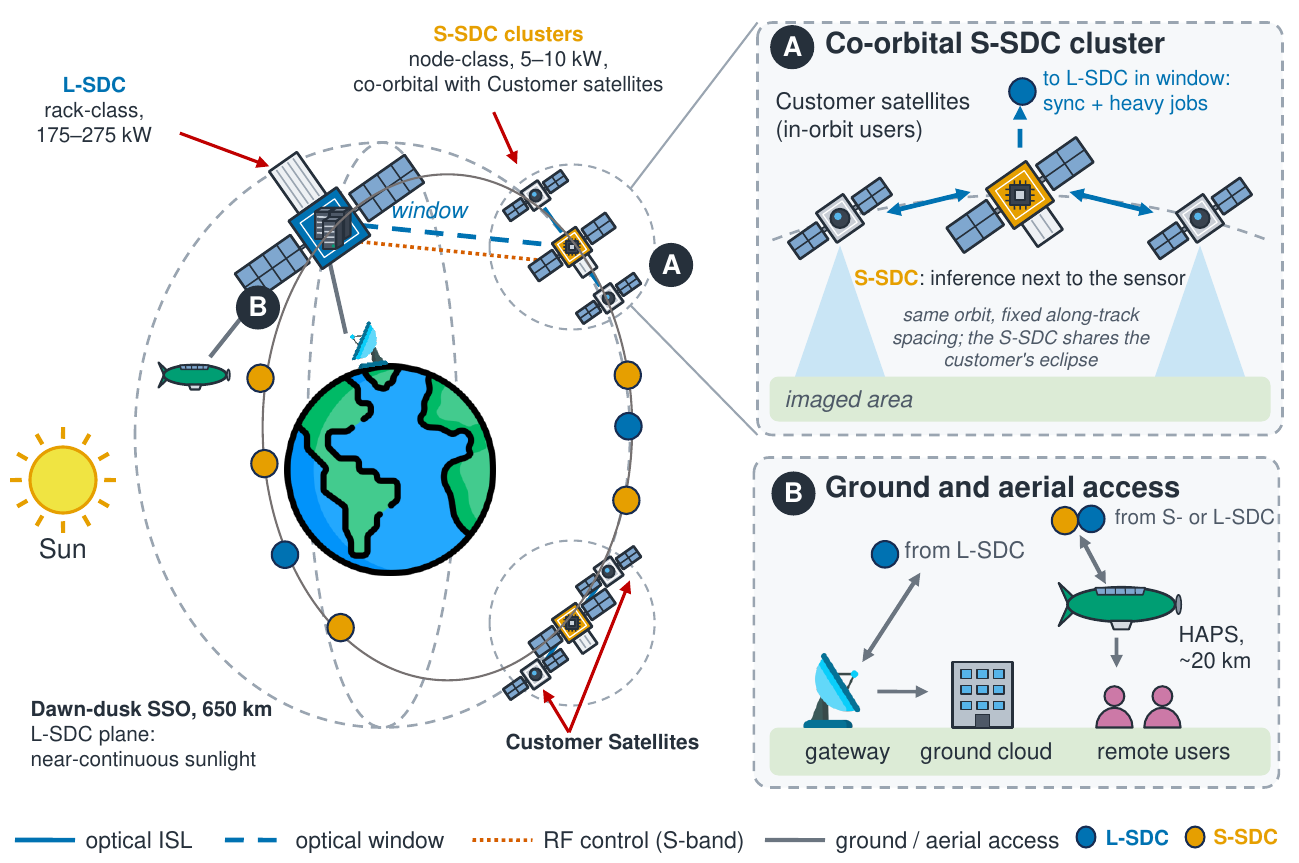}}
\caption{Proposed HSDC with large nodes (L-SDCs, blue) in a dawn-dusk orbit and small nodes (S-SDCs, orange) in their customers' orbits. (A) Co-orbital cluster. (B) Ground and aerial access.}\vspace*{-5pt}
\label{fig:hsdc}
\end{figure*}

As shown in Figure~\ref{fig:hsdc}, L-SDC nodes operate in a dawn-dusk SSO and form a shared backbone for computing, storage, and model management, whereas S-SDCs operate in the orbits of the customer spacecraft they serve, providing in-situ computing near where data are generated.

\subsection{Example Large and Small SDC Nodes}
We consider an L-SDC a rack-class node built around a single NVL72-type accelerator rack (Table~\ref{tab:gpu}). Power conditioning, thermal transport, pointing, and communications add about 25\% to the rack's computing power, so each L-SDC must generate and reject roughly 175--275~kW \cite{turyshev2026odc}. At 350~K and an emissivity of 0.92, this requires at least 225--350~m$^2$ of radiator area. Environmental heat absorption increases this area by 25\% to 67\% \cite{turyshev2026odc}. At 325~W/m$^2$ at the beginning of life, the solar array requires 540--850~m$^2$. This is about 24\% of the 1,361~W/m$^2$ solar flux, for 30\% cell efficiency~\cite{nasa2026soa} and 20\% losses from cell packing, temperature, and power conversion. Degradation adds about 18\% over the mission~\cite{turyshev2026odc}. A dedicated space-based solar power (SBSP) feeder is an option for dense L-SDC clusters~\cite{alhassan2026constellation}.

An S-SDC has a 5--10~kW power budget, comparable to one eight-GPU NVIDIA HGX H100 server. It flies in its customers' orbits and shares their eclipses, which last up to 35~min per 98-min orbit at 650~km. Its array must therefore supply about 1.6 times the load in sunlight \cite{turyshev2026odc}. For a 10-kW node, this corresponds to about 50~m$^2$ of array area and about 20~kWh of battery capacity at a 30\% depth of discharge (DoD), a conservative value for LEO batteries~\cite{turyshev2026odc}.

An L-SDC's power budget is 20 to 55 times that of an S-SDC, and its radiator and array scale accordingly. Hundreds of square meters of deployed surface would add drag and collision risk to a customer constellation, so an L-SDC suits a dedicated orbit chosen for power availability. A 10-kW S-SDC, in contrast, fits within current large LEO buses. Radiation also separates the two roles. Proton tests of a commercial tensor processing unit showed no hard failures well beyond the expected five-year dose in a shielded SSO, and the silent-data-corruption rate was acceptable for inference but remains an open issue for training \cite{aguera2025space}. We therefore assign inference to S-SDCs and training to L-SDCs, where checkpointing and redundancy can be shared across accelerators.

\subsection{Users, System Composition, and Placement}

An HSDC serves in-orbit users (customer spacecraft), ground users through gateways, and aerial users through HAPS. Figure~\ref{fig:hsdc} shows one L-SDC plane and two customer planes, each with an S-SDC cluster (inset A), and the ground and aerial access paths (inset B).

We place L-SDCs in a dawn-dusk SSO at 650~km altitude with an inclination of 97.98$^\circ$, following Google's Suncatcher~\cite{aguera2025space} and the 600~km dawn-dusk case analyzed by~\cite{turyshev2026odc}. In this orbit, the beta angle stays near 80$^\circ$, above the 65.2$^\circ$ needed to stay out of Earth's shadow. Each L-SDC is therefore in sunlight almost all the time and needs only a small safe-mode battery. The cost is coverage: the orbit follows the line between day and night, so it passes over low- and mid-latitude users only near dawn and dusk.

We place each S-SDC in the orbital plane of the customer constellation it serves, at the same altitude and with a fixed along-track spacing (inset A). The inter-satellite link (ISL) range is therefore constant, pointing is almost static, and contact is continuous. With customers a few hundred kilometers apart, commercial data-center transceivers can close the link, and the delay stays below 1~ms~\cite{aguera2025space}. In sparser planes, such as the eight-satellite plane in our evaluation, ranges reach a few thousand kilometers and need a long-range optical terminal, but the delay stays below 10~ms. An S-SDC and the customers within its reach form a cluster. Several clusters in the same plane can form a ring for load sharing and backup.

Customer planes differ from the dawn-dusk plane in right ascension, so S-SDCs meet L-SDCs in link windows near the plane crossings, around $80^\circ$ latitude for a 10:30 plane. At equal altitude, the phase between two nodes is locked. It must be small enough to keep them in link range, but not zero, or they would meet at the same point at every crossing.

\subsection{Coordination Between S-SDC and L-SDC Nodes}

Coordination is based on a functional split of the computing tasks. The L-SDC hosts centralized functions, including the model repository, training and fine-tuning, large-batch jobs, long-term storage, fleet-level scheduling and DT, and the primary path to the ground. The S-SDC hosts local functions, including inference on raw sensor data, filtering and compression, job admission from its cluster, and scheduling within its power and thermal limits. Time-critical jobs that fit within the S-SDC budget are processed locally, while larger jobs are forwarded to an L-SDC in the next link window. An S-SDC entering eclipse can likewise hand over tasks to a sunlit L-SDC~\cite{liu2024phoenix}.

\begin{figure}
\centerline{\includegraphics[width=18.5pc]{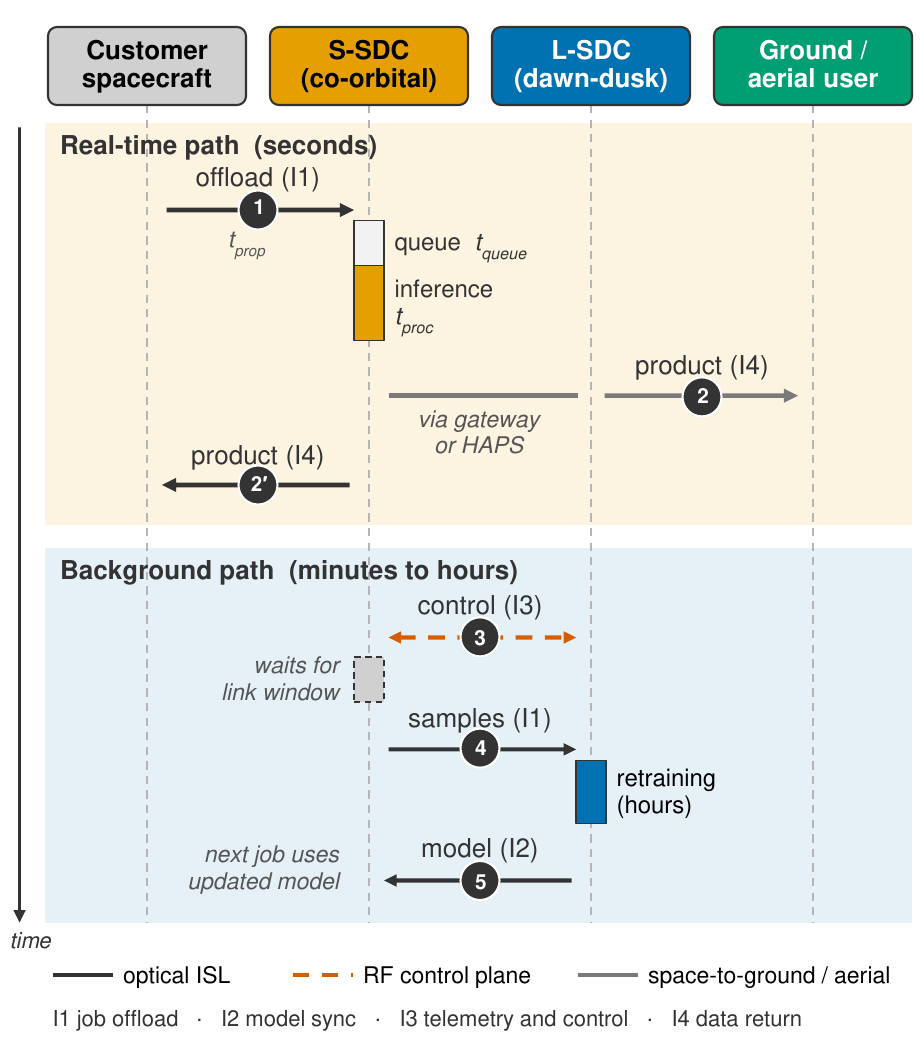}}
\caption{AI-in-the-loop workflow of an HSDC. Steps 1--5 and interfaces I1--I4 are described in the text.}\vspace*{-5pt}
\label{fig:workflow}
\end{figure}

We define four logical interfaces for this coordination, as shown in Figure~\ref{fig:workflow}. Interface I1 (job offloading and data ingestion) transfers raw data from customer spacecraft to their S-SDCs and samples or large jobs from S-SDCs to an L-SDC. Interface I2 (model synchronization) delivers model updates from the L-SDC to its S-SDCs during link windows. Interface I3 (telemetry and orchestration) transmits health status, energy state, and scheduling decisions among all nodes and the ground segment over the RF control plane. Interface I4 (data return) delivers results to customer spacecraft, to ground users via gateways, and to aerial users via HAPS.

Figure~\ref{fig:workflow} depicts the AI-in-the-loop workflow for the representative service considered, namely processing high-resolution EO imagery, although the same workflow applies to other in-orbit services. A scene is the unit of work, defined as a ground area image captured by EO satellites. A customer spacecraft offloads its data to the S-SDC over I1 (step 1). The S-SDC runs inference and returns a compact product over I4, either to the ground via a gateway or HAPS (step 2) or to an in-orbit user (step 2$'$), while exchanging telemetry and control over I3 (step 3). In the background, the S-SDC uploads selected hard samples to an L-SDC in the next link window (step 4). The L-SDC retrains the model and returns it over I2 (step 5). Steps 1, 2, and 2$'$ form the real-time path, whose latency is the sum of propagation, queuing, and processing delays. Since propagation is at most about 18~ms even at the line-of-sight limit, processing and queuing dominate this latency. Steps 4 and 5 form the background path, which waits for L-SDC link windows and is limited by link bandwidth rather than latency, since retraining takes hours rather than seconds.

Despite its limited availability, the L-SDC provides greater computing capacity. We consider a GPU-second as the workload of one H100 GPU for one second; therefore, a 100 GPU-second job requires about 12 seconds on an eight-GPU S-SDC. Beyond a compute threshold, the S-SDC saturates, and the L-SDC path becomes faster. This crossover determines the S-SDC admission threshold and forms the main trade-off between the two node classes, which we quantify alongside latency in the evaluation.

\subsection{Interconnection and Space-to-Ground Access}

The HSDC interconnects its nodes through two link tiers. Tier~1 is an S-band RF control plane, which we assume operates at 1--100~Mb/s; it carries I3 traffic and small I4 products and does not require precise pointing. Tier~2 is an optical data plane at 1550~nm, over which commercial ISLs, such as those of Starlink, already achieve about 100~Gb/s at ranges up to 5,400~km~\cite{aguera2025space}. On these links, pointing and acquisition, rather than transmit power, limit performance, and each link window loses part of its duration to acquisition.

L-SDCs serve as the primary gateways between the HSDC and the ground because the dawn-dusk orbit crosses the polar regions on every revolution, so a small number of high-latitude gateways can reach each L-SDC once per orbit. Optical downlinks from LEO have reached 100--200~Gb/s in demonstrations, but sustained rates are much lower because visibility, weather, pointing, and scheduling reduce usable contact time~\cite{turyshev2026odc}. The HSDC therefore returns compact products rather than raw data. An S-SDC can also reach ground users directly, though a LEO node remains visible from a ground location for only a few minutes per pass. Two additional paths complement the gateways. Optical relays in geostationary orbit (GEO) provide near-continuous visibility, and we use them as a baseline for our evaluation. HAPS operate above most weather and can store results between passes for users without terrestrial network infrastructure.

In summary, the proposed architecture defines the role of each node class, the orbit in which each class operates, and the interfaces that carry each type of traffic. 

\section{PRELIMINARY EVALUATION RESULTS}\label{sec:eval}

We evaluate the HSDC with a time-stepped simulation that propagates orbits using Kepler’s equations, including Earth’s oblateness, and models links, queues, and processors at 10-s steps.

\subsection{Setup}

We consider a customer plane of eight EO satellites at 650 km, with a 10:30 local time at the ascending node. A dawn-dusk plane hosts up to eight L-SDCs. Up to four S-SDCs are positioned in the customer plane, each midway between two neighboring EO satellites. These satellites are 45$^\circ$ apart; each S-SDC is 2,740 km from its two customers. Three gateways at Svalbard, Troll, and Winnipeg provide ground access. Svalbard and Troll are the two commercial polar sites that see a sun-synchronous satellite on nearly every orbit, while Winnipeg is a mid-latitude user site. Each EO satellite generates 2 GB scenes while in sunlight, averaging 400 per day and up to 1,600 in a stress case, and transmits them over optical ISLs at 10 Gb/s. We use two workloads from \cite{liu2024phoenix}, ship detection and wildfire segmentation, which they timed on the NVIDIA Jetson AGX Orin, a low-power onboard AI computer. Based on the specification sheets, scaled to an H100 with a speed ratio of 15, they require 0.2 and 3.4 GPU-seconds per scene. Workloads produce 20 MB products, and we report the segmentation workload unless noted.

Three configurations receive identical scene streams: ground-only, where raw scenes are downlinked at 1.2 Gb/s \cite{eoportal_wv3}; L-SDC-only, where each EO satellite sends its scenes to an L-SDC during link windows; and the HSDC, where each EO satellite sends its scenes to its S-SDC. Each run spans 48 hours, including a 24-hour measurement window, and is repeated across five random sequences of scene arrivals.

\subsection{Node Sizing and Placement}

\begin{figure}
\centerline{\includegraphics[width=18.5pc]{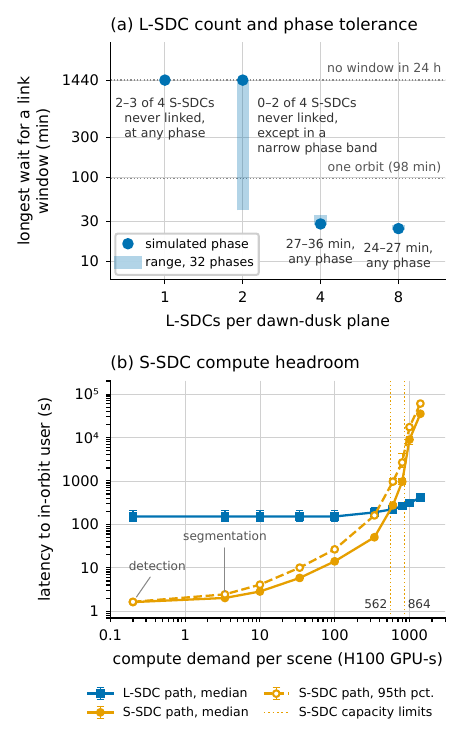}}
\caption{Node sizing and placement. (a) Longest wait for an L-SDC link window vs. L-SDC count across 32 initial phases of the L-SDC plane. (b) Latency to the in-orbit user vs. compute demand per scene at 400 scenes per satellite per day.}\vspace*{-5pt}
\label{fig:sizing}
\end{figure}

Figure~\ref{fig:sizing}(a) shows how many L-SDCs a plane needs. For each L-SDC count, the dot indicates the worst-case wait for the worst-placed S-SDC within a link window. The band shows how this wait varies across 32 starting positions, or phases, of the L-SDCs along their orbit. Because both planes share the same altitude and period, the phase between an S-SDC and an L-SDC remains constant. The two nodes therefore come within link range only if their phase difference is below about 50$^\circ$. With one or two L-SDCs, at least half of the S-SDCs never meet one. With four L-SDCs spaced 90$^\circ$ apart, every S-SDC obtains two windows per orbit at any phase, each lasting 8 to 21 min, and the longest wait is 27 to 36 min. Eight L-SDCs shorten this wait by less than 10 min. We therefore recommend four L-SDCs per plane, with a phase offset of a few degrees to keep the nodes apart at the crossings.

Figure~\ref{fig:sizing}(b) plots latency to the in-orbit user against compute demand per scene for the HSDC and L-SDC-only paths. For both workloads, the eight-GPU S-SDC is lightly loaded, and latency stays near 2~s, mostly from transferring the raw scene over the 10-Gb/s link. The S-SDC path remains faster up to about 450 GPU-seconds per scene. Beyond that, S-SDC queues build in sunlight, when all scenes arrive (about 65\% of each orbit), and the 72-GPU L-SDC becomes faster. The S-SDC saturates at 562 GPU-seconds per scene over the sunlit arc and 864 GPU-seconds per scene for the daily mean.

\subsection{Service Performance}

\begin{figure}
\centerline{\includegraphics[width=18.5pc]{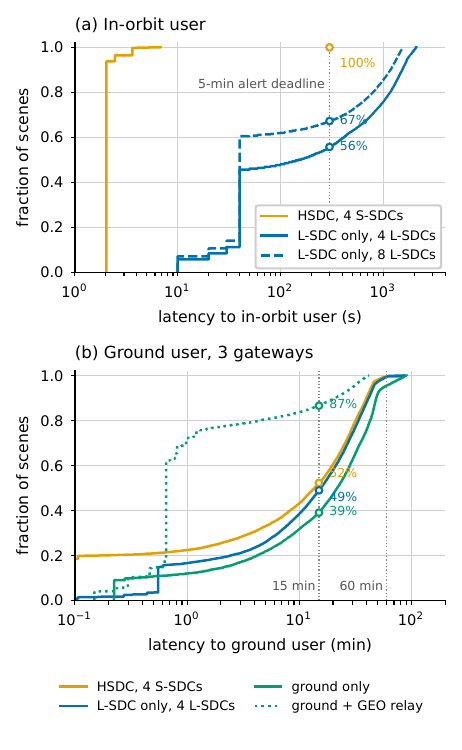}}
\caption{Latency distributions at 400 scenes per satellite per day for (a) in-orbit users and (b) ground users. Markers give the fraction of scenes delivered within each deadline.}\vspace*{-5pt}
\label{fig:cdf}
\end{figure}

Figure~\ref{fig:cdf} shows latency distributions for 400 scenes per satellite per day. Each curve shows the fraction of scenes delivered within a given time; curves farther to the left indicate faster service. The HSDC returns products to the in-orbit user in about 2s (1.6s transfer), with every product delivered within 7s.

The L-SDC-only curve has two parts, appearing as a discontinuity. About 46\% of scenes are captured while a link window is open and are served within 40 s (30 s for link acquisition, typical of commercial laser terminals \cite{bhattacharjee2023lisl}). Other scenes wait for the next window, so the 95th percentile across all scenes reaches 29 min; just 56\% meet a 5-min alert deadline. This gap stems from placement, not speed: the L-SDC processes scenes faster than the S-SDC, but it can be reached only during windows.

Ground users must wait for satellites to pass over a gateway. Median latency is 14 min for the HSDC, 16 min for L-SDC-only, and 22 min for ground-only; 52\%, 49\%, and 39\% of products arrive within 15 min, respectively. In the L-SDC-only configuration, the number of L-SDCs determines the total number of scenes delivered. Four L-SDCs imply full delivery, two imply half, and one implies a quarter, since other EO satellites never meet L-SDCs. The HSDC delivers every scene completely, as its S-SDCs send directly to the ground. The ground-only configuration saturates at 1,600 scenes per satellite per day when gateway passes are too short to download all raw data. A GEO optical relay at 1.8 Gb/s~\cite{esa_edrs} removes this limit, delivering 87\% of products within 15 min, but in our model it serves no other users. The HSDC therefore provides ground users with only a modest latency gain, and none compared with a relay. Its benefit is lower downlink volume: 20 MB per product vs 2 GB per scene for images, whereas the relay carries every raw scene. Even when it serves only our satellites, it is busy 63\% of the time, handling 1,600 scenes per satellite per day. A relay shared with other missions would soon run out of capacity.

\subsection{Energy by Role and Scale}
Assuming a 35\% peak power draw on accelerators when idle, energy use is dominated by idle power. The L-SDCs consume about 1,470~kWh per day, and S-SDCs about 60 kWh. Together, four L-SDCs and four S-SDCs consume $\sim$6,100 kWh per day, with 96\% from the L-SDCs. This cost scales with nodes, not with load. Since segmentation workloads add only 0.5 Wh per scene on S-SDCs and 1.5 Wh on L-SDCs, the L-SDC energy cost is fixed, while the S-SDC energy cost grows with customer clusters. During eclipses, an S-SDC draws 1.4 kWh per orbit at this load and needs a 4.7 kWh battery at 30\% DoC, compared with $\sim$20 kWh for a 10-kW node that runs at full power.

Our model excludes certain details. It does not forward jobs from busy S-SDCs to L-SDCs, retrain L-SDCs, or account for weather and atmospheric drag. Additionally, the H100 speed is scaled from Jetson timings using specification sheets, not measurements. These effects change the absolute numbers but not the configuration rankings. Forwarding would only improve the HSDC; weather affects all ground links; drag is small over two days; and Figure~\ref{fig:sizing}(b) covers a wide range of compute demand.

\section{DESIGN GUIDANCE AND RESEARCH DIRECTIONS}\label{sec:guidance}

Our results lead to five design rules for an HSDC. First, small nodes should be placed with their customers. An S-SDC in the customer plane returns products within a few seconds, whereas a larger node in another plane is reachable only during link windows. Denser S-SDCs placed closer to in-orbit, or NTN users allow shorter, higher-rate links and fewer users per node, which can shorten the AI computation loop further. Second, a dawn-dusk plane needs about four L-SDCs, with a small phase offset from the S-SDCs, so that no S-SDC waits more than about 36 min for a link window. Third, an S-SDC should be sized for the peak arrival rate in sunlight rather than the daily mean, with jobs above a few hundred GPU-seconds forwarded to an L-SDC. Fourth, the HSDC should send compact products rather than raw data to the ground. Ground latency is mainly determined by gateway passes, while downlink volume determines whether space-to-ground links keep up. Fifth, L-SDC energy should be planned as a shared backbone cost, while S-SDC batteries can be sized for the actual load in eclipse rather than for full power.

Several research directions follow from the parts that our model simplifies. The first is the admission and forwarding policy between node classes, which must decide in real time where each job runs as loads, link windows, and battery states change. The second is model synchronization over short link windows, where the L-SDC must retrain and distribute models to S-SDCs in windows of only 8 to 21 min. The third is reliable computing on commercial accelerators, since radiation effects are acceptable for inference but remain an open issue for training on the L-SDC. The fourth is energy supply for L-SDCs at hundreds of kilowatts, through larger arrays, better radiators, or dedicated SBSP feeders. Fifth, HAPS could store results between passes and extend service to aerial users in regions without gateways. Finally, an HSDC shared by multiple operators will require common interfaces and fleet-level orchestration, for which DTs of nodes and links are a natural tool.

In summary, HSDC offers a practical path from single orbital computers to networked SDCs serving ground, aerial, and in-orbit users.

\section{ACKNOWLEDGMENTS}
This work was supported by the Natural Sciences and Engineering Research Council of Canada (NSERC) under Grant RGPIN-2022-03364, and by Research Manitoba.

\def\refname{REFERENCES}

\vspace*{-8pt}

\begin{IEEEbiography}{Peng Hu}{\,}is an Associate Professor in the Department of Electrical and Computer Engineering at the University of Manitoba and an Adjunct Associate Professor in the Cheriton School of Computer Science at the University of Waterloo, Canada. He serves as Secretary of the IEEE ComSoc's Satellite and Space Communications TWG, Associate Editor for the \textit{IEEE Transactions on Aerospace and Electronic Systems} and West Area Chair of Region 7. His research interests include non-terrestrial networks, space sustainability, edge AI and autonomous networking. Contact him at peng.hu@umanitoba.ca.
\end{IEEEbiography}

\begin{IEEEbiography}{Mustafa Alhassan}{\,} is a Ph.D. student in the Department of Electrical and Computer Engineering at the University of Manitoba, Winnipeg, MB R3T 5V6, Canada. His research interests include digital twins-driven resilience of satellite networks, space-based computing and power, and non-terrestrial networks. He is a Member of IEEE. Contact him at alhass11@myumanitoba.ca.
\end{IEEEbiography}

\begin{IEEEbiography}{Amjad Iqbal} {\,} is a Postdoctoral Research Fellow in the Department of Electrical and Computer Engineering at the University of Manitoba, Canada. His research interests include 5G/6G and non-terrestrial networks, integrated sensing and communication, and reconfigurable intelligent surfaces. Contact him at amjad.iqbal@umanitoba.ca.
\end{IEEEbiography}

\end{document}